\documentclass[12pt]{article}
\usepackage{epsfig}
\usepackage{amssymb}
\usepackage{amsfonts}
\usepackage{subfigure}
\usepackage{color}
\def\beq{\begin{equation}}
\def\eeq{\end{equation}}
\def\bea{\begin{eqnarray}}
\def\eea{\end{eqnarray}}
\def\ksl{\hbox{\hbox{${k}$}}\kern-1.9mm{\hbox{${/}$}}}

\newcommand{\text}{\rm}

\def\lsim{\raise0.3ex\hbox{$\;<$\kern-0.75em\raise-1.1ex\hbox{$\sim\;$}}} 

\def\gsim{\raise0.3ex\hbox{$\;>$\kern-0.75em\raise-1.1ex\hbox{$\sim\;$}}}

\begin{document}

\begin{center} 
{\bf \large  Sum rules and spectral density flow in QCD and \\ in superconformal theories
\footnote{Presented at \emph{QCD@Work2014}, 16-19 June 2014, Giovinazzo (Bari-Italy)}
}
\vspace{0.2cm}
{\bf \Large } 

\vspace{1.5cm}
{\bf Antonio Costantini, Luigi Delle Rose and Mirko Serino}
\vspace{1cm}

{\it Dipartimento di Matematica e Fisica "Ennio De Giorgi", 
Universit\`{a} del Salento and \\ INFN-Lecce, Via Arnesano, 73100 Lecce, Italy\footnote{antonio.costantini@le.infn.it, luigi.dellerose@le.infn.it, mirko.serino@le.infn.it}}\\

\vspace{.5cm}
\begin{abstract} 
We discuss the signature of the anomalous breaking of the superconformal symmetry in $\mathcal{N}=1$ super Yang Mills theory and its manifestation in the form of anomaly poles.  
Moreover, we describe the massive deformations of the $\mathcal{N}=1$ theory and the spectral densities of the corresponding anomaly form factors.  
These are characterized by spectral densities which flow with the mass deformation and turn the continuum contributions from the two-particle cuts of the intermediate states into poles, with a single sum rule satisfied by each component. 
The poles can be interpreted as signaling the exchange of a composite axion/dilaton/dilatino (ADD) multiplet in the effective Lagrangian. We conclude that global anomalous currents characterized by a single flow in the perturbative picture always predict the existence of composite interpolating fields.
\end{abstract}
\end{center}

\newpage

\section{Introduction}
The question whether the Higgs is a fundamental scalar or a composite particle remains unanswered at this time, but most likely it will be answered in the near future at the LHC, in the next scheduled run. In the meanwhile, there are some puzzling features of the Standard Model which suggest that the Higgs could be accompanied by an extra 
composite state, which is the Nambu-Goldstone mode of a broken conformal symmetry. According to recent analyses, this state could be the physical manifestation of the trace anomaly in the Standard Model, in analogy to the pion, which is interpolated by the chiral current and by the corresponding $\langle AVV \rangle$ (axial-vector/vector/vector) interaction in QCD. In this case, as for the case of chiral symmetry, the description of the dynamics involving this extra state would require a nonlinear realization, with an extra potential which is necessary in order to lift its mass by a small finite amount. Evidence in favour of this scenario comes directly from perturbation theory, and brings us quite far to 
propose that all the anomalies, conformal or chiral, are mediated, according to perturbation theory, by effective scalar or pseudoscalar degrees of freedom. In particular,  we have shown rigorously that in the Standard Model all the gauge invariant sectors of the theory carry this specific signature. We are going briefly to review these aspects, which emerge from studies of anomalous correlators in QED \cite{Giannotti:2008cv,Armillis:2009im,Armillis:2009pq}, QCD \cite{ Armillis:2010qk} and in the electroweak theory \cite{Coriano:2011zk}, now extended to supersymmetric gauge theories \cite{Coriano:2014gja}. 

As in the $\langle AVV \rangle$ case,  this composite state should be identified with the anomaly 
pole of the related anomaly correlator (the $\langle TVV \rangle$ diagram, with $T$ the energy momentum tensor (EMT)) \cite{Coriano:2012nm}. The anomaly pole identified as a possible effective scalar in gravity in \cite{Giannotti:2008cv} has been shown to appear as a manifestation of an effective state interpolated by the dilatation current of the Standard Model in \cite{Coriano:2012nm}. Considerations of this nature brings us to the conclusion that the effective massless Nambu-Goldstone modes which should appear as a result of the existence of global anomalies, should be looked for in specific perturbative form factors under special kinematical limits. As discussed in \cite{Coriano:2014gja}, these are the limits which guarantee the infrared coupling of this pole. 

\section{The superconformal anomalous correlators}
We turn our attention to the effective action of the superconformal (the Ferrara-Zumino, FZ \cite{Ferrara:1974pz}) multiplet, where chiral and conformal anomalies share similar signatures, being part of the same multiplet. \\
We have shown in perturbation theory that the anomaly of the FZ multiplet is associated with the exchange of three composite states in the 1PI superconformal anomaly action. 
They are identified with the anomaly poles present in the effective action, extracted from a supersymmetric correlator $\langle \mathcal{J V V} \rangle$ in $\mathcal N=1$ Super Yang-Mills theory containing the superconformal hypercurrent $\mathcal{J}$ and two vector currents $\mathcal V$, and correspond to the dilaton, the dilatino and the axion. 
This exchange is identified by a direct analysis of the anomalous correlators in perturbation theory or by the study of the flow of their spectral densities under massive deformations.
The flow describes a 1-parameter family of spectral densities - one family for each component of the correlator - which satisfy mass independent sum rules, and are, therefore, independent of the superpotential. This behaviour turns a dispersive cut of the spectral density $\rho(s,m^2)$ into a pole (i.e. a $\delta(s)$ contribution) as the deformation parameter $m$ goes to zero. 
Moreover, denoting with $k^2$ the momentum square of the anomaly vertex, each of the spectral densities induces on the corresponding form factor a $1/k^2$ behaviour also at large $k^2$, as a consequence of the sum rule. \\
The three states associated to the three anomaly poles mentioned above are described - in the perturbative picture - by the exchange of two collinear particles. These correspond to a fermion/antifermion pair in the axion case, a fermion/antifermion pair and a pair of scalar particles in the dilaton case, and a collinear scalar/fermion pair for the dilatino. 
\\
This pattern appears to be general in the context of anomalies \cite{Giannotti:2008cv,Armillis:2009pq, Armillis:2010qk, Armillis:2009im}, and unique in the case of supersymmetry. In fact, in a supersymmetric theory anomaly correlators have a single pole in each component of the anomaly multiplet, a single spectral flow and a single sum rule, proving the existence of a one-to-one correspondence between anomalies and poles in these correlators \cite{Coriano:2014gja}. 
\\
Here we present the one-loop perturbative analysis of the three one-particle irreducible correlators, each of them built with a single current insertion, which can be extracted, in the component formalism, from the $\langle \mathcal{J V V} \rangle$. These are $\Gamma_{(R)}$, $\Gamma_{(S)}$ and $\Gamma_{(T)}$ defined as
\begin{eqnarray}
\label{RSTCorrelators}
\delta^{ab} \, \Gamma_{(R)}^{\mu\alpha\beta}(p,q) &\equiv& \langle R^{\mu}(k)\, A^{a \, \alpha}(p) \, A^{b \, \beta}(q) \rangle \qquad \langle RVV \rangle \,, \nonumber \\
\delta^{ab} \, \Gamma_{(S) \, A\dot B}^{\mu\alpha}(p,q) &\equiv& \langle S^{\mu}_A (k) \, A^{a \, \alpha}(p) \, \bar \lambda^b_{\dot B}(q) \rangle \qquad \langle SVF \rangle \,, \nonumber \\
\delta^{ab} \, \Gamma_{(T)}^{\mu\nu\alpha\beta}(p,q) &\equiv& \langle T^{\mu\nu}(k) \, A^{a \, \alpha}(p) \, A^{b \, \beta}(q)\rangle \qquad \langle TVV \rangle  \,,
\end{eqnarray}
with $k = p+q$, and where $R$, $S$ and $T$ are respectively the $R$-current, the supersymmetric current and the energy-momentum tensor, while $A^{a \, \alpha}$ and $\lambda^a_A$ are the gauge and the gaugino fields. \\
We show the results in the case of a massive chiral multiplet running in the loops. The contributions coming from the gauge vector superfield share a similar behaviour and can be found in \cite{Coriano:2014gja}. The presence of a mass parameter turns out to be extremely useful in order to discuss the general behaviour of the spectral densities away from the conformal point. The explicit perturbative computation gives
\begin{eqnarray}
\label{RChiralOSMassive}
\Gamma_{(R)}^{\mu\alpha\beta}(p,q) &=&  i \frac{g^2 \, T(R)}{12 \pi^2} \, \Phi_1(k^2,m^2) \, \frac{k^\mu}{k^2} \epsilon[p, q, \alpha ,\beta]  \,, \\
\label{SChiralOSMassive}
\Gamma^{\mu\alpha}_{(S)}(p,q) &=&  i \frac{g^2 T(R)}{6 \pi^2 \, k^2} \, \Phi_1(k^2,m^2) \, s_1^{\mu\alpha}
+ i \frac{g^2 T(R)}{64 \pi^2}  \, \Phi_2(k^2,m^2) \, s_2^{\mu\alpha} \,, \\
\label{TChiralOSMassive}
\Gamma_{(T)}^{\mu\nu\alpha\beta}(p,q) &=& \frac{g^2 \, T(R)}{24 \pi^2 \, k^2} \, \Phi_1(k^2,m^2) \, t_{1S}^{\mu\nu\alpha\beta}(p,q) + \frac{g^2 \, T(R)}{16 \pi^2} \, \Phi_2(k^2,m^2) \, t_{2S}^{\mu\nu\alpha\beta}(p,q) \,, 
\end{eqnarray}
where the tensor structures are defined as
\begin{eqnarray}
s_1^{\mu\alpha} &=&   \sigma^{\mu \nu} k_\nu \, \sigma^\rho k_\rho \,  \bar \sigma^{\alpha \beta} p_\beta \,,\nonumber \\ 
s_2^{\mu\alpha} &=&  2 p_\beta \, \sigma^{\alpha \beta} \sigma^\mu \,, 	\nonumber \\
t_{1S}^{\mu\nu\alpha\beta} &=&  (\eta^{\mu\nu} k^2 - k^\mu k^\nu) u^{\alpha\beta}(p,q)\,, \nonumber \\
t_{2S}^{\mu\nu\alpha\beta} &=&  (p^\mu q^\nu + p^\nu q^\mu) \eta^{\alpha\beta} + p \cdot q (\eta^{\alpha\nu} \eta^{\beta\mu} + \eta^{\alpha\mu} \eta^{\beta\nu}) - \eta^{\mu\nu} u^{\alpha \beta}(p,q) - (\eta^{\beta\nu}p^\mu + \eta^{\beta\mu}p^\nu)q^\alpha \nonumber \\
&-&   (\eta^{\alpha\nu}q^\mu + \eta^{\alpha\mu}q^\nu)p^\beta\,,
\end{eqnarray}
with $u^{\alpha\beta}(p,q) = \eta^{\alpha\beta} p \cdot q - p^{\beta} q^\alpha$.
The two form factors are
\begin{eqnarray}
\Phi_1(k^2,m^2) &=& - 1 - 2\, m^2 \, \mathcal C_0(k^2,m^2) \,, \nonumber \\
\Phi_2(k^2,m^2) &=& 1 - \mathcal B_0(0,m^2) + \mathcal B_0(k^2,m^2) + 2 m^2  \mathcal C_0(k^2,m^2) \,.
\label{exp1}
\end{eqnarray}
where $\mathcal B_0$ and $\mathcal C_0$ are the standard two- and three-point scalar integrals. It is important to observe that the pole contributions in the correlators come from the anomalous structures only, which are multiplied by the anomalous form factor $\Phi_1$. 

\section{The sum rule}
As pointed out long ago in the literature on the chiral anomaly \cite{Dolgov:1971ri}, its perturbative signature is in the appearance of a massless pole 
(an anomaly pole) in the spectrum of the $\langle AVV \rangle$ diagram. The pole is present, in perturbation theory, only in a specific kinematic configuration, namely at zero fermion mass and with on-shell vector lines. The intermediate state which is exchanged in the effective action and which is mediated by the $\langle AVV \rangle$ diagram, is characterized by two massless and collinear fermions moving on the light cone. It is rather compelling to interpret the appearance of this intermediate configuration - within the obvious limitations of the perturbative picture - as signalling the possible exchange of a bound state in the quantum effective action. 

In a phenomenological context, what gives a far broader significance to the appearance of this kinematic singularity, which accompanies any perturbative anomaly is the existence of a sum rule for the spectral density $\rho(s,m^2)$ of the anomaly form factor. Generically, it is given in the form 
\begin{equation}
\label{sr}
\frac{1}{\pi} \int_0^{\infty}\rho(s, m^2) ds =  f,
\end{equation}
with the constant $f$ independent of any mass (or other) parameter which characterizes the thresholds or the strengths of the resonant states eventually present in the integration region $(s>0)$.
It should be stressed that sum rules formulated for the study of the structure of the resonances, i.e. their strengths and masses, as in the QCD case, involve a parameterization of the resonant behaviour 
of $\rho(s,m)$ at low $s$ values, via a phenomenological approach, with the inclusion of the asymptotic behaviour of the correlator, amenable to perturbation theory, for larger $s$. For this significant interplay between the infrared (IR) and the ultraviolet (UV) regions, the term {\em duality} is indeed quite appropriate to qualify the implications of a given sum rule.

The spectral density associated with the anomaly form factor of the superconfomal anomaly correlators, which is given by
\begin{equation}
 \chi(k^2, m^2)\equiv \Phi_1(k^2,m^2)/k^2, 
\end{equation}
can be computed from its discontinuity and reads as
\begin{equation}
\label{spectralrho}
{\rho}_\chi(s,m^2) = \frac{1}{2 i} \textrm{Disc} \, \chi(s,m^2) = \frac{2 \pi m^2}{s^2}\log\left(\frac{1 + \sqrt{\tau(s,m^2)}}{1-\sqrt{\tau(s,m^2)}}\right)\theta(s- 4 m^2), \qquad \tau(s,m^2) = \sqrt{1 -4m^2/s}.
\end{equation}
A specific feature of the spectral density of the superconformal anomalies, but also of the chiral and the conformal ones, is that the pole is introduced in the spectrum in a specific kinematical limit, as a degeneracy of the two-particle 
cut when any second scale (for instance the mass) is sent to zero 
\begin{equation}
\lim_{m\to 0} \frac{1}{\pi} \rho_\chi(s,m^2) =  \delta(s).
\end{equation}
The property of the {\em cut turning into a pole} is peculiar to finite (non superconvergent) sum rules (here $\rho_\chi$ has been normalized to $f=1$). It is related to a spectral density which is normalized by the sum rule just like an ordinary weighted distribution, and whose support is located at the edge of the allowed phase space ($s=0$) as the conformal deformation turns to zero. This allows to single out a unique interpolating state out of all the possible exchanges permitted in the continuum, i.e. for $s> 4 m^2$, as the theory flows towards its conformal/superconformal point.\\
Elaborating on Eq. (\ref{sr}), one can show that 
the effect of the anomaly is, in general, related to the behaviour of the spectral density at any values of $s$, although, in some kinematical limits, it is the region around the light cone ($s\sim 0$) which dominates the sum rule, and amounts to a resonant contribution.
In fact, the combination of the scaling behaviour of the corresponding form factor $F(Q^2)$ (equivalently of its density $\rho$) with the requirement of integrability of the spectral density, essentially fix $f$ to be a constant and the sum rule (\ref{sr}) to be saturated by a single massless resonance. Obviously, a superconvergent sum rule, obtained for $f=0$, would not share this behaviour. At the same time, {\em the absence of subtractions} in the dispersion relations guarantees the significance of the sum rule, being this independent of any ultraviolet cutoff. 

It is quite straightforward to show that Eq. (\ref{sr}) is a constraint on  asymptotic behaviour of the related form factor.
The proof is obtained by observing that the dispersion relation for a form factor in the spacelike region ($Q^2=-k^2> 0$)
\begin{equation}
F(Q^2,m^2)=\frac{1}{\pi}\int_0^{\infty} ds\frac{ \rho(s,m^2) }{s + Q^2} \,,
\end{equation}
once we expand the denominator in $Q^2$ as $\frac{1}{s + Q^2}=\frac{1}{Q^2} - \frac{1}{Q^2} s\frac{1}{Q^2} +\ldots$ and make use of Eq. (\ref{sr}), induces the following asymptotic behaviour on $F(Q^2,m^2)$
\begin{equation}
\lim_{Q^2\to \infty} Q^2 F(Q^2, m^2)=f.
\label{asym}
\end{equation}
The $F\sim f/Q^2$ behaviour at large $Q^2$, with $f$ independent of $m$, shows the pole dominance of $F$ for $Q^2 \rightarrow \infty$. The UV/IR conspiracy of the anomaly,  discussed in \cite{Armillis:2009im, Armillis:2009sm,Armillis:2009pq}, is in the reappearance of the pole contribution at very large value of the invariant $Q^2$, even for a nonzero mass $m$. 
This point is quite subtle, since the flows of the spectral densities with $m$ show the decoupling of the anomaly pole for a nonzero mass. Here, the term {\em decoupling} will be used to refer to the non resonant behaviour of $\rho$.
Therefore, the presence of a $1/Q^2$ term in the anomaly form factors is a property of the entire flow which a) converges to a localized massless state (i.e. $\rho(s)\sim\delta(s)$) as 
$m\to 0$, while  
b) the presence of a non vanishing sum rule guarantees the validity of the asymptotic constraint illustrated in Eq. (\ref{asym}). Notice that although for conformal deformations driven by a single mass parameter the independence of the asymptotic value $f$ on $m$ is a simple consequence of the scaling behaviour of $F(Q^2,m^2)$, it holds quite generally even for a completely off-shell kinematics \cite{Giannotti:2008cv}.

In summary, for a generic supersymmetric $\mathcal{N}=1$ theory, the two basic features of the anomalous behaviour of a certain form factor responsible for chiral or conformal anomalies are: \,\,1)  the existence of a spectral flow which turns a dispersive cut into a pole as $m$ goes to zero and 2) the existence of a sum rule which relates the asymptotic behaviour of the anomaly form factor to the strength of the pole resonance.\\
 In a supersymmetric theory this correspondence is unique, since the only poles present in the explicit expressions are those 
 of the anomaly form factors. This feature is shared also by the $\langle AVV \rangle$ in non supersymmetric theories, where one can identify 
 a single pole in the related form factor, a single sum rule and a single spectral density flow. In the $\langle TVV \rangle$ diagram, for a general field theory, instead, this feature is absent. 
 The appearance of extra poles in the form factors of the traceless parts of this second correlators leaves unanswered the question about the physical meaning of these additional singularities \cite{Giannotti:2008cv,Armillis:2009pq, Armillis:2010qk, Armillis:2009im}. \\
In order to clarify how the cancellation of the extra poles occurs for the supersymmetric $\langle TVV \rangle$, we consider the non-anomalous form factor $\Phi_2$ of the traceless part of the correlator in a general theory, with $N_f$ Weyl fermions, $N_s$ complex scalars and $N_A$ gauge fields. We work, for simplicity, in the massless limit. In this case $\Phi_2$, which is affected by pole terms can be written in the form
\begin{eqnarray}
\label{F2total}
\Phi_2(k^2) &=& \frac{N_f}{2} \Phi_2^{(f)}(k^2) + N_s \, \Phi_2^{(s)}(k^2) + N_A \,  \Phi_2^{(A)}(k^2) \nonumber \\
&=& \frac{g^2}{144 \pi^2 \, k^2} \left[ -  \frac{N_f}{2} T(R_f) + N_s \, \frac{T(R_s)}{2} + N_A \, \frac{T(A)}{2} \right] \,, 
\end{eqnarray}
where the fermions give a negative contribution with respect to scalar and gauge fields. 
If we turn to a $\mathcal N=1$ Yang-Mills gauge theory, we need to consider in the anomaly diagrams the virtual exchanges both of a chiral and of a vector supermultiplet. In the first case the multiplet is built out of one Weyl fermion and one complex scalar, therefore in Eq.(\ref{F2total}) we have $N_f = 1, N_s = 1, N_A = 0$ with $T(R_f) = T(R_s)$. With this matter content, the form factor is set to vanish. \\
For a vector multiplet, on the other end, we have one vector field and one Weyl fermion, all belonging to the adjoint representation and then we obtain $N_f=1, N_s=0, N_A=1$ with $T(R_f) = T(A)$. Even in this case all the contributions in the $\Phi_2$ form factor sum up to zero. It is then clear that the cancellation of the extra poles in the $\langle TVV \rangle$ is a specific tract of supersymmetric Yang Mills theories, due to their matter content, not shared by an ordinary gauge theory. A corollary of this is that in a supersymmetric theory we have just one spectral flow driven by the deformation parameter $m$, accompanied by one sum rule for the entire deformation.

\section{Conclusions}
We have presented  additional evidence that anomaly poles are the signature of the anomalies in the perturbative anomaly action of these theories, extending former studies \cite{Giannotti:2008cv,Armillis:2010qk,Armillis:2009im}. For global anomalies it is expected that the massless states identified by the pole contributions can be promoted to new composite degrees of freedom by some non perturbative dynamics, as for the chiral anomaly and the pion.\\
In supersymmetric theories, we have shown that the connection between anomalies, poles and sum rules for anomaly vertices are one to one.
The $1/ k^2$ feature of the anomaly form factors has been investigated in connection with the scaling properties of their spectral densities and with the finite (non zero) sum rule. We have seen that the anomalous behaviour emerges from the $s\sim 0$ region of the spectral density of a given form factor and covers, therefore, 
the entire light-cone surface. The resonant behaviour at $s=0$ is present, as we have shown, also at very high momentum.\\
We have focused our attention on the perturbative correlators which are responsible for the generation of the superconformal anomaly, and which allow to identify some composite 
states in the effective action, interpolating between the currents and the on-shell final states. They correspond to a dilaton, a dilatino and an axion. \\
Following this pattern, it is then natural to ask if global anomalies are always connected to the generation of effective degrees of freedom, and hence to compositeness, as indicated by the poles of the effective action. These results are valid for all the anomalies characterized by a single flow, in particular for all the chiral currents affected by global anomalies. From this perspective, also the Peccei-Quinn current should induce as an interpolating  state a composite axion rather an elementary one, being our argument generic to anomalous global currents.

\end{document}